\documentclass[prb,showpacs,twocolumn,floatfix,superscriptaddress]{revtex4}
\usepackage{epsfig}

\begin{document}

\title{Multichannel effects in Rashba quantum wires}
\author{M. M. Gelabert}
\affiliation{Departament de F\'{\i}sica, Universitat de les Illes Balears,
E-07122  Palma de Mallorca, Spain}
\author{Lloren\c{c} Serra}
\affiliation{Departament de F\'{\i}sica, Universitat de les Illes Balears,
E-07122  Palma de Mallorca, Spain}
\affiliation{ Institut de F\'{\i}sica Interdisciplinar i de Sistemes Complexos
IFISC (CSIC-UIB), E-07122 Palma de Mallorca, Spain.}
\author{David S\'anchez}
\affiliation{Departament de F\'{\i}sica, Universitat de les Illes Balears,
E-07122  Palma de Mallorca, Spain}
\author{Rosa L\'opez}
\affiliation{Departament de F\'{\i}sica, Universitat de les Illes Balears,
E-07122  Palma de Mallorca, Spain}

\date{January 26, 2010}

\begin{abstract}
We investigate intersubband mixing effects in multichannel
quantum wires in the presence of Rashba spin-orbit coupling and
attached to two terminals. When the contacts
are ferromagnetic and their magnetization direction is perpendicular
to the Rashba field, the spin-transistor current is expected
to depend in a oscillatory way on the Rashba coupling strength
due to spin coherent oscillations of the travelling electrons.
Nevertheless, we find that the presence of many propagating
modes strongly influences the spin precession effect,
leading to (i) a quenching
of the oscillations and (ii) strongly irregular curves
for high values of the Rashba coupling. We also observe that in the case
of leads' magnetization parallel to the Rashba field,
the conductance departs from a uniform value as the
Rashba strength increases.
We also discuss the Rashba interaction induced
current polarization effects when the contacts are not magnetic
and investigate how this mechanism is affected by the presence of
several propagating channels.
\end{abstract}
\maketitle

\section{Introduction}

Since the discovery of the giant magnetoresistance effect,\cite{Fert88,Grunberg89}
research in spintronics has been developing at a fast pace.
An important requirement for practical applications of
this novel technology is the generation,
control and manipulation of spin-polarized currents
preferably using electric fields only.\cite{fab07}
Spin-orbit interactions in semiconductor materials
are promising tools to achieve that goal. In particular,
the Rashba interaction,\cite{Rashba60} a type of spin-orbit coupling
that originates from a lack of inversion symmetry in
semiconductor heterostructures
(such as InAs or GaAs), has been experimentally shown to possess
a high degree of tunability using gate contacts.\cite{nit97}

Since the spin-orbit interaction couples the electron momentum
and its spin, the Rashba field behaves as an effective magnetic field that is
responsible for spin coherent oscillations,
which can be exploited in spintronics.
Based on this property, Datta and Das suggested a
spin field-effect transistor.\cite{Dat90}
It consists of a one-dimensional ballistic
channel sandwiched by two ferromagnetic contacts.
Their proposal relies on the control of the current along the channel
using the Rashba interaction via a third terminal (the gate)
and the relative orientation of the leads' magnetizations.
The length of the channel and the intensity of the Rashba strength determine
the flow of the current. Realization of the spin transistor was hindered by some limitations, such as
the mismatch problem (which results in poor injection of spin-polarized
current between a ferromagnet and a semiconductor)\cite{schmidt} and
the idealization of ballistic transport.\cite{schliemann} However,
recent experiments on quasi-two dimensional structures\cite{Koo09},
already discussed in Refs.~\onlinecite{zai,baby},
have overcome these obstacles and have obtained
a behavior which looks similar to the spin transistor effect.

In reality, strictly one-dimensional channels are hard to fabricate
and one must deal mostly with {\em quasi}-one dimensional systems
containing many propagating channels.
Confinement in the transversal direction is accomplished with
potentials leading to subband spacings often smaller than a few~meV,
the order of magnitude of the Fermi energy in low-dimensional systems.
As a consequence, multiple subbands are populated and channel
mixing effects become relevant in many situations. In fact,
the Rashba interaction itself includes an intersubband mixing term
which couples adjacent subbands with opposite spins.
This coupling has been recently demonstrated to give rise
to strongly modulated conductance curves,\cite{she04,zhan05,san06,lop07} especially close
to the onset of higher-energy plateaus, due to Fano interference\cite{fano}
between propagating waves and Rashba induced localized levels.\cite{san06}
In the presence of in-plane magnetic fields,
Rashba coupling induced intersubband mixing effects
are shown\cite{ser05} to reduce the visibility of anomalous conductance steps,\cite{per04}
and to produce transmission asymmetric lineshapes even in purely one-dimensional
systems.\cite{san08}

In this paper, we analyze the role of intersubband coupling effects in multichannel
quantum wires. Our model consists of a quantum wire with a localized Rashba
spin-orbit interaction coupled to ferromagnetic leads with magnetization perpendicular
to the direction of the Rashba field.
We find that the Rashba intersubband coupling term modifies the spin precession
effect in a dramatic way. Typically, one finds a few oscillation cycles in the conductance
curves before arriving at a strongly irregular domain at high values of the Rashba
parameter in which case the intersubband coupling produces an effective
randomization of the injected spins independent of the relative orientation
of the leads' magnetization. Therefore, our results point out
a serious limitation of the spin transistor performance, even in the ideal cases of perfect
spin injection and fully ballistic propagation.

On the other hand, Rashba interaction has lately deserved much attention
as a generation procedure of spin-polarized currents. Several methods
have been proposed in different setups
(see Refs.~\onlinecite{kis01,gov02,ion03,usa04,gov04,kho04,ya05,eto05,ohe05,sil06,%
cum06,zha07,per07,liu07,lu07,scheid,ape08,aha08,zha08,deb09,gel09,ent10},
although the list is by no means exhaustive).
We here consider a simple system: a Rashba quantum wire
attached to two nonmagnetic leads.
We find that the Rashba interaction can
produce a highly polarized electric current and that the effect is purely due
to interchannel coupling. For quantum waveguides supporting a single
propagating mode, the polarization effect vanishes.\cite{bul02,zha05,kis05}
Since the Rashba interaction
is localized, we calculate the generated polarization as a function of the
interface smoothness and show that the highest values of the polarization
are obtained when the transition between the regions with and without
spin-orbit interaction is abrupt.

In Sec.~II we discuss the physical system and establish
the theoretical model to calculate the linear conductance.
Section~III is devoted to the numerical results when
the contacts are ferromagnetic. The spin polarization
effect in the case of normal contacts is analyzed
in Sec.~IV. Finally, Sec.~V contains our conclusions.

\section{Physical system and model}

\begin{figure}[t]
\centerline{
\epsfig{file=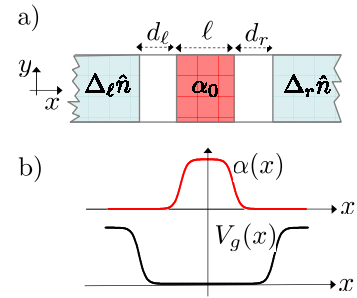,angle=0,width=0.25\textwidth,clip}
}
\caption{(Color online)
Sketch of the physical system (a) and of the spatial variation
of Rashba intensity $\alpha(x)$ and gate potential $V_g(x)$ (b).
}
\label{fig1}
\end{figure}

We consider a quasi-one dimensional system (a quantum wire)
with a localized Rashba interaction (the Rashba dot)
coupled to semi-infinite leads.
Figure \ref{fig1} shows a sketch of the physical system.
Transport occurs along the $x$ direction.
We characterize the Rashba dot as a small region of length $\ell$
with strong spin-orbit coupling with strength $\alpha_0$.
The spin polarization in the leads
is described using the Stoner model for itinerant ferromagnets.
Due to exchange interaction among the electrons, the electronic
bands in the asymptotic regions become spin split with a splitting phenomenologically
given by an effective field $\Delta_0$, which we take as a parameter.
This approximation is good at low temperatures (lower than the Curie temperature)
and for electron densities large enough so that strong correlations
can be safely neglected.\cite{auerbach}
Denoting the Stoner field in left and right regions by
$\Delta_\ell$ and $\Delta_r$, respectively,
the parallel configuration is described by
$\Delta_\ell=\Delta_r=\Delta_0$ while the antiparallel
corresponds to $\Delta_\ell=-\Delta_r=\Delta_0$.
In addition, we assume that a local gate potential $V_g(x)$ is
aligning the potential bottom of the successive regions.
This way we remove unwanted conductance modifications due to the potential
mismatches,\cite{schmidt}
thus focussing on the properties induced purely by the spin-orbit coupling.

The system Hamiltonian reads
\begin{eqnarray}
{\cal H} &=&
-\frac{\hbar^2}{2m}\left(
\frac{d^2}{dx^2}+\frac{d^2}{dy^2}
\right)
+\frac{1}{2}m\omega_0^2y^2\nonumber\\
&+& V_g(x)+\Delta(x)\, \hat{n}\cdot\vec{\sigma}
+{\cal H}_R\; .
\end{eqnarray}
The confinement
along the direction $y$, perpendicular to the current,
is taken as parabolic with oscillator
frequency $\omega_0$, which defines the length $\ell_0=\sqrt{\hbar/m\omega_0}$.
The inhomogeneous Rashba coupling ${\cal H}_R$
is given by
\begin{eqnarray}
{\cal H}_R &\equiv&
{\cal H}_R^{(1)} +
{\cal H}_R^{(2)} \nonumber\\
&=&
\frac{\alpha(x)}{\hbar}p_y\sigma_x+
\left(
-\frac{\alpha(x)}{\hbar} p_x
+\frac{i}{2}\,\alpha'(x)
\right)
\sigma_y\; ,
\label{eqR}
\end{eqnarray}
where, as usual, spin is represented by the vector of Pauli matrices $\vec\sigma$
while $p_x$ and $p_y$ are the Cartesian components of the electron's linear
momentum. The Rashba intensity $\alpha(x)$ varies smoothly taking a constant
value $\alpha_0$ inside the Rashba dot and vanishing elsewhere.
The term proportional to $p_x$ is responsible for spin precession
of an injected electron.\cite{Dat90} The intersubband coupling
term proportional to $p_y$ couples adjacent subbands with opposite spins.
Finally, the term with the derivative $\alpha'(x)$ is added in Eq.\ (\ref{eqR}) to ensure
the Hermitian character of the Hamiltonian.

As mentioned above, the Stoner field $\Delta(x)$ is constant in the left and right asymptotic regions
($\Delta_{\ell,r}$) and it smoothly vanishes at distances $d_{\ell,r}$ towards the
left and right of the Rashba dot.
These are assumed large enough such that all
evanescent states at the interface vanish before reaching the leads. The gate
potential aligning the band bottom of the different regions is taken as
$V_g(x)=|\Delta(x)|$. An equivalent choice but localized to the Rashba dot
would be
$V_g(x)=|\Delta(x)|-\Delta_0$.
All spatial transitions in
$\alpha(x)$ and $\Delta(x)$ are described using Fermi-like type functions
characterized by a small diffusivity $a$.\cite{Fermi} In general,
$a$ is assumed to be small enough, although we shall also
discuss below the dependence with this parameter in some cases.

For a given energy $E$ the electron wave function fulfills
Schr\"odinger's equation
\begin{equation}
({\cal H}-E)\Psi=0\; ,
\label{eqS}
\end{equation}
with the appropriate boundary conditions. Our method of solution
combines discretization of the longitudinal variable $x$ in a
uniform grid with a basis expansion in transverse eigenfunctions
$\phi_n(y)$ and in eigenspinors $\chi_s(\eta)$ along
a direction given by a unitary vector $\hat{n}$,
\begin{equation}
\Psi=
\sum_{s=\pm}
\sum_{n=0}^{\infty}
\psi_{ns}(x)\,\phi_n(y)\,\chi_s(\eta)\; ,
\label{exp}
\end{equation}
where $s=\pm$ is the spin quantum number while $\eta=\uparrow,\downarrow$
denotes the twofold spin discrete variable.
In terms of the polar and azimuthal angles $(\theta,\phi)$
corresponding to the spin quantization axis $\hat{n}$ we
can write
\begin{equation}
\chi_+\equiv
\left(
\begin{array}{c}
\cos\left(\frac{\theta}{2}\right)\\
\rule{0cm}{0.5cm}
\sin\left(\frac{\theta}{2}\right)\,e^{i\phi}
\end{array}
\right)
\;;\;
\chi_-\equiv
\left(
\begin{array}{c}
\sin\left(\frac{\theta}{2}\right)\\
\rule{0cm}{0.5cm}
-\cos\left(\frac{\theta}{2}\right)\,e^{i\phi}
\end{array}
\right)\; .
\end{equation}
The transverse eigenfunctions are the solutions of the
harmonic 1D oscillator
\begin{equation}
\left(
-\frac{\hbar^2}{2m}
\frac{d^2}{dy^2}
+\frac{1}{2}m\omega_0^2y^2
\right)\,\phi_n(y)
=
\varepsilon_n\,\phi_n(y)\; ,
\end{equation}
with
\begin{equation}
\varepsilon_n=\left(n+\frac{1}{2}\right)\hbar\omega_0\;;\quad n=0,1,\dots\; .
\end{equation}

Projecting Eq.\ (\ref{eqS}) onto the basis we obtain the equations
for the unknown {\em channel amplitudes} $\psi_{ns}(x)$
\begin{eqnarray}
-\frac{\hbar^2}{2m} \psi_{ns}^{''}(x)
&+&\left(\rule{0cm}{0.5cm}
V_g(x)+s\Delta(x)+\varepsilon_n-E
\right)\psi_{ns}(x)\nonumber\\
&+&
\sum_{n's'}{
\langle ns|{\cal H}_R | n's'\rangle\,
 \psi_{n's'}(x)}=0\; .
\label{ccm}
\end{eqnarray}
Notice that the Rashba interaction is the only source of
interchannel coupling since, in general, the matrix element
$\langle ns|{\cal H}_R | n's'\rangle$ will be non diagonal.
Using the separation in two spin-orbit contributions
introduced in Eq.\ (\ref{eqR}) we can write
\begin{eqnarray}
\label{eq9}
\langle ns|{\cal H}_R^{(1)} | n's'\rangle
&=&
\frac{\alpha(x)}{\hbar}\langle n|p_y|n'\rangle \langle s |\sigma_x|s'\rangle\; ,\\
\label{eq10}
\langle ns|{\cal H}_R^{(2)} | n's'\rangle
&=&
\left(
-\frac{\alpha(x)}{\hbar} p_x
+
\frac{i}{2}\alpha'(x)
\right)\delta_{nn'}
\langle s |\sigma_y|s'\rangle\; .\nonumber\\
\end{eqnarray}
Equations (\ref{eq9}) and (\ref{eq10}) clearly show that,
in general, both ${\cal H}_R^{(1)}$ and ${\cal H}_R^{(2)}$
couple channels with opposite spins through the
matrix elements $\langle s|\sigma_{x}|s'\rangle$
and $\langle s|\sigma_{y}|s'\rangle$.
Of course, if the spin quantization axis $\hat{n}$
is chosen along the $x$ or $y$ axis then either
$\langle s|\sigma_{x}|s'\rangle$ or
$\langle s|\sigma_{y}|s'\rangle$ become diagonal.
Regarding the coupling between transverse modes, we notice that
${\cal H}_R^{(2)}$ is always diagonal ($\delta_{nn'}$) while
${\cal H}_R^{(1)}$ is connecting modes differing in one subband index
($n'=n\pm 1$) through the oscillator matrix element
$\langle n|p_y|n'\rangle$.

If we neglect ${\cal H}_R^{(1)}$
as in strict one-dimensional systems, Eq.\ (\ref{ccm})
involves a single mode $n$. If, in addition, the spin axis
is chosen along $y$ then the two spin modes uncouple and no spin
oscillation is allowed; in other directions ($x$ or $z$) a rigid spin
precession should be expected if all the contribution
between parenthesis in Eq.\ (\ref{eq10}) is assumed constant.
This precession is the underlying working mechanism of the
Datta-Das spin transistor.\cite{Dat90} Below we investigate the solution
of Eq.\ (\ref{ccm}) in the general case in order to analyze the robustness
of the {\em spin precession} scenario when ${\cal H}_R^{(1)}$
is included and when space inhomogeneity in $\alpha(x)$ is also
taken into account. The Appendix contains the details of the
employed numerical method to compute the transmission
$t_{n's',ns}$, i.e., the probability amplitude from a given left incident mode $ns$
to the right mode $n's'$. Then, using the scattering approach
the linear-response conductance is given by,
\begin{equation}
G=
G_0
\sum_{ns,n's'}
\left|
t_{n's',ns}
\right|^2\; ,
\end{equation}
where $G_0=e^2/h$ is the conductance quantum.
For later discussion on the polarization of the transmitted
current we also define the {\em polarized}
conductance $G_p$,
\begin{equation}
\label{eqgp}
G_p =
G_0
\sum_{ns,n's'}
s'\left|
t_{n's',ns}\right|\; ,
\end{equation}
and the relative polarization $p$
$(-1\le p\le 1)$,
\begin{equation}\label{eqp}
p =\frac{G_p}{G}\; .
\end{equation}
We shall pay special attention to the multichannel case considering
energies $E$ in Eq.\ (\ref{ccm}) such that
up to 10 propagating modes are active in the leads.

\begin{figure}[t]
\centerline{
\epsfig{file=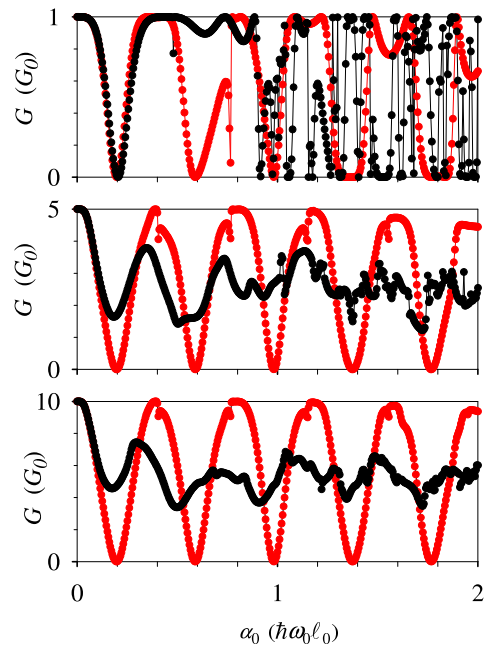,angle=0,width=0.37\textwidth,clip}
}
\caption{(Color online)
Conductance as a function of Rashba coupling intensity.
Black corresponds to the
complete Rashba interaction while
grey (red color) to the neglect of ${\cal H}_R^{(1)}$.
The leads are spin-polarized along $x$.
Upper, intermediate and lower panel correspond to
$N_p=1$, 5 and 10 propagating modes, respectively.  We take the parameters
$\ell=8\ell_0$, $E=N_p\hbar\omega_0$, $\Delta_\ell=\Delta_r=10\hbar\omega_0$,
$d_\ell=d_r=10 \ell_0$, $a=0.1\ell_0$.}
\label{fig2}
\end{figure}

\begin{figure}[t]
\centerline{
\epsfig{file=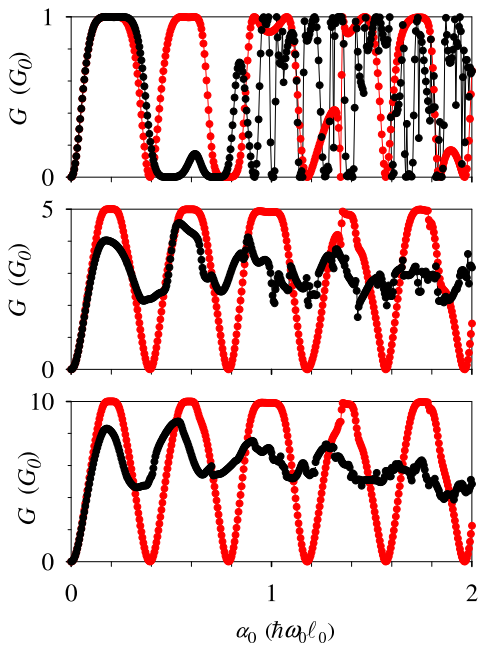,angle=0,width=0.37\textwidth,clip}
}
\caption{(Color online)
Same as Fig.\ \ref{fig2} for polarized leads along $x$ but
in antiparallel orientations, i.e.,
$\Delta_\ell=10\hbar\omega_0$ and
$\Delta_r=-10\hbar\omega_0$.}
\label{fig3}
\end{figure}

\section{Results for spin polarized leads}

Figure \ref{fig2} shows the results for polarized leads oriented
along $x$. When ${\cal H}_R^{(1)}$ is neglected the conductance
for 5 and 10 propagating modes displays an almost sinusoidal
behavior with only minor distortions. These deviations, which
are enhanced in
the single mode case, can be attributed to the quantum interference
with the Rashba dot.\cite{san06}
The present results confirm, therefore, the precession scenario
mentioned above
but only
when the number of modes is large enough
and interband coupling is neglected.
Quite remarkably,
however, this scenario is not robust with the inclusion
of ${\cal H}_R^{(1)}$. When the full Rashba interaction is
considered only for small values of $\alpha_0$ the conductance behaves in a
regular way. Very rapidly as $\alpha_0$ increases $G$
fluctuates in a staggered way that resembles the conductance fluctuations
of disordered systems.
The mean value, in units of $G_0$, is $\approx 0.5 N_p$, with $N_p$ the number of
active channels, while the amplitude of the
fluctuation decreases when $N_p$ increases.

The existence of the first conductance minimum has been
clearly seen in the experiments of Ref.\ \onlinecite{Koo09}. Our
results are in agreement with this experiment, but they also predict
that successive maxima and minima are heavily
distorted or even fully washed out. It is also worth noticing that
the first conductance minimum for the black dots occurs at a slightly
lower value of $\alpha_0$ than that of the grey (red color)
data, indicating that the minima
$\alpha_{\rm min}$ are somewhat
contracted with
respect to the simple prediction from
the Rashba dot length:  $2m\ell\alpha_{\rm min}=n\pi\hbar^2$, with
$n=1,2,\dots$ (red symbols).

Figure \ref{fig3} contains the results for polarized leads
along $x$ but in antiparallel directions. In this case, when
$\alpha_0\approx0$ the conductance vanishes due to the
spin valve effect. As $\alpha_0$ increases, however, the conductance
rises and the spin valve effect is effectively destroyed by
the presence of the Rashba dot. For big enough values the
system behaves similarly to the case of parallel polarized leads
(Fig.\ \ref{fig2}), displaying irregular oscillations around a mean
value $\approx N_p/2$. For strong spin-orbit
couplings and high number of modes no clear distinction
between parallel and antiparallel orientations is then to be expected.
This is a consequence of the strong subband mixing. In fact,
if ${\cal H}_R^{(1)}$ is neglected (red symbols) there is a
full correspondence between the conductance nodes of the parallel
geometry with the maxima of the antiparallel one; as could
expected from the simplified rigid precession scenario.

The above results are not modified if other values of
$\Delta_{\ell,r}$ are used, provided they are large enough
to ensure full polarization of the leads. The same is true
for distances $d_{\ell,r}$. They should be large enough to
allow the decay of evanescent states at the interfaces with
the Rashba dot and at the points where Stoner fields are
switched on.

\begin{figure}[t]
\centerline{
\epsfig{file=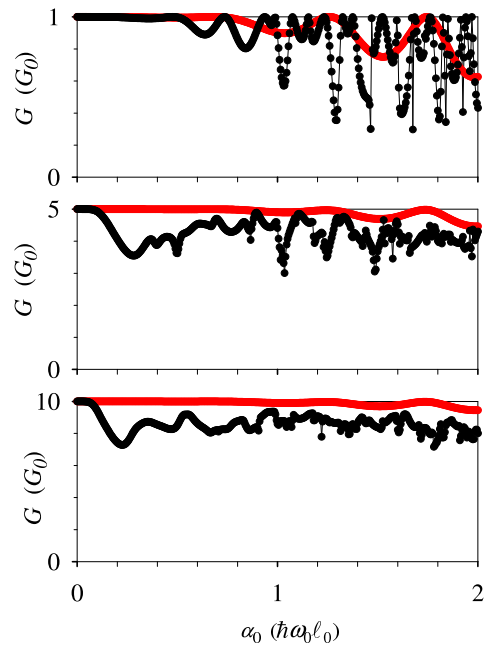,angle=0,width=0.37\textwidth,clip}
}
\caption{(Color online)
Same as Fig.\ \ref{fig2} for parallel polarized leads along $y$.}
\label{fig4}
\end{figure}

\begin{figure}[t]
\centerline{
\epsfig{file=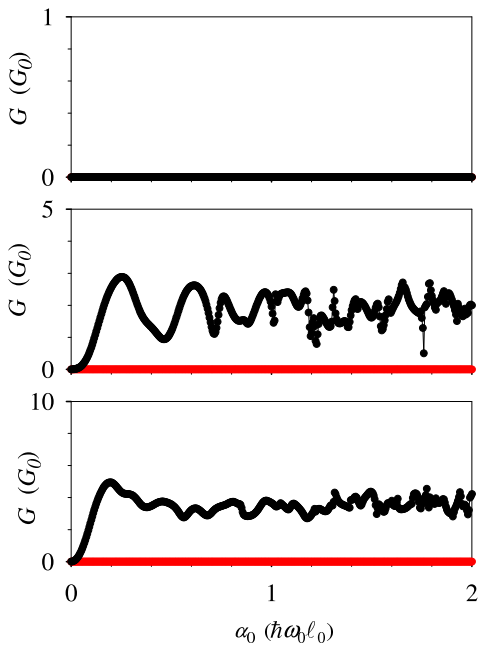,angle=0,width=0.37\textwidth,clip}
}
\caption{(Color online)
Same as Fig.\ \ref{fig2} for polarized leads along $y$
in antiparallel orientation.}
\label{fig5}
\end{figure}

We consider next polarized leads along $y$ and $z$; that is, in
directions that are perpendicular to the quantum wire.
For $z$ polarizations the results are very similar to the $x$ ones already
discussed and thus will not be shown. Figures
\ref{fig4} and \ref{fig5}
contain the results for $y$-polarized parallel and antiparallel
leads. A first conspicuous difference with the results of
Figs.\
\ref{fig2} and \ref{fig3}
is that the grey symbols (red color) do not display wide
sinusoidal oscillations. The conductance when ${\cal H}_R^{(1)}$
is neglected is actually maximal for the parallel case and
stays rather constant with some
small oscillations at large $\alpha$'s that disappear when the number of channels
increases.
On the other hand, $G$
vanishes for the antiparallel orientation.
We understand this spin-valve behavior as a complete absence of spin precession,
resulting from the fact that ${\cal H}_R$
is spin diagonal in this approximation [cf.\ Eq.\ (\ref{eq10})].

Including ${\cal H}_R^{(1)}$ in the $y$-polarized geometry again
yields qualitative modifications of the linear conductance (black
symbols in Figs.\
\ref{fig4} and \ref{fig5}).
Except for the antiparallel one-channel case,
$G$ shows staggering behavior at large $\alpha_0$'s, quite
similarly to the $x$-polarized results. On average, the conductance is somewhat
reduced from the maximal value in the parallel case (Fig.\ \ref{fig4}) and,
remarkably, takes a finite value in the antiparallel distribution (Fig.\ \ref{fig5}).
For $\alpha_0\sim 0.2\hbar\omega_0\ell_0$ the antiparallel
conductance has already reached
a value close to $N_p/2$ and to the eventual saturation value.
The Rashba coupling is thus quite effective in allowing
transmission by flipping spins of the polarized incoming
electrons towards the opposite spin orientation of the
outgoing ones. The single channel limit
(upper panel of Fig.\ \ref{fig5})
is obviously an exception since even the black symbols
vanish in this case. This is
easily understood noticing that the incident
$ns=0+$ mode couples in the Rashba dot
with modes $1-, 2+, \dots$, but not with $0-$,
which is the only propagating mode in the right lead.
Therefore, no conduction is possible under
this conditions.

Experimentally, the absence of conductance oscillation
in the parallel $y$-oriented configuration has been
confirmed.\cite{Koo09} Our results reproduce that behavior
(Fig.\ \ref{fig4}) and they also suggest the antiparallel
$y$ orientation (Fig.\ \ref{fig5}) as an interesting
configuration for a spin-orbit-controlled device.
Indeed, the initial rise of conductance in the multichannel
case, interpreted above as a Rashba-induced destruction of the spin valve,
could be used as the conducting (ON) state of the device.
One should check, however, that the
evolution of $G(\alpha_0)$ from zero to the higher values
remains smooth
for increasing numbers of propagating channels.
The present results do not elucidate this point but they
seem to indicate that for $N_p=10$ propagating modes the initial
rise of $G(\alpha_0)$
occurs more rapidly than for $N_p=5$.
In a future work we shall treat
the continuum case, having an infinite number of transverse states,
using a different approach from the present one.

The results shown above are not much modified if the interfaces
with the Stoner fields
at distaces $d_\ell$ and $d_r$ to the left and right of the Rashba
dot, respectively (See Fig.\ \ref{fig1}), are smoothed by increasing the
corresponding Fermi-function parameter.\cite{Fermi}
This confirms
that the conductance modifications are an effect of the Rashba dot,
and not of the Stoner field interfaces.
Indeed, the more diffuse the interface, the more reflectionless and
thus more ideal is the description of the contact.
In the next section we shall discuss the case of nonpolarized leads ($\Delta_0=0$),
but we have also calculated some cases of partial polarization by decreasing $\Delta_0$
when both $s=+$ and $-$ transverse states are active, although their number is not
perfectly balanced. We have found that the conductance is qualitatively similar
to the fully polarized case, with irregular behaviour at large values of $\alpha_0$.

\section{Rashba polarizers}

It has been recently pointed out\cite{per07,gel09}
that a Rashba dot can act as a current polarizer in such a way that when a
non polarized current enters the dot from the left, the transmitted current
to the right may
attain an important degree of spin polarization in $y$ direction.
For this to occur, it has been shown that at least two propagating
modes of opposite spin must interfere.\cite{per07,gel09} In wires with parabolic transverse
confinement this
means that the energy should at least exceed $1.5\hbar\omega_0$ such that
the four modes $\{0+,0-,1+,1-\}$ are active and the interference occurs
in subsets $\{0+,1-\}$ and $\{0-,1+\}$.
The resulting spin polarization is
very sensitive to the energy
(see Fig.\ 3 of Ref.\ \onlinecite{gel09}) and a large enhancement of the polarization
$p$, Eq.\ (\ref{eqp}) is obtained when the energy is such that
a Fano-type resonance with a quasibound state from a higher evanescent
band is formed. This type of resonances which lead to the Fano-Rashba effect was
investigated in Ref.\ \onlinecite{san06}.
The polarization of the transmitted current is zero if, instead of $y$,
other direction for the quantization axis are chosen.

The preference for the transverse $y$ direction in polarization
is an example of {\em chirality} induced by the Rashba interaction.
This is possible even with a time-reversal invariant Hamiltonian
like Eq.\ (\ref{eqR}) because our boundary condition (left incidence)
is not time reversal invariant. Indeed, if we consider
the time reversed boundary condition, i.e.,
incidence from
the right, the current transmitted to the left is polarized in the opposite
direction.
The superposition of both solutions completely restores the
symmetry without any preferred spin direction.
The reversal of the polarization for the right-to-left
transmission can be seen as a peculiar behavior of Rashba polarizers
that makes them {\em fragile} in the presence of magnetic barriers
like those of Sec. III. Indeed, one could naively think that when the
Rashba dot acts as
a current polarizer the left-to-right transmission
with $y$-magnetized leads
should be very high
in parallel configuration and very low
in antiparallel configuration.
This is not the case, however, because of multiple
backwards and forwards
reflections with their associated
inversions of $p$ (see lower panels
of Figs.\ \ref{fig4} and \ref{fig5}).

\begin{figure}[t]
\centerline{
\epsfig{file=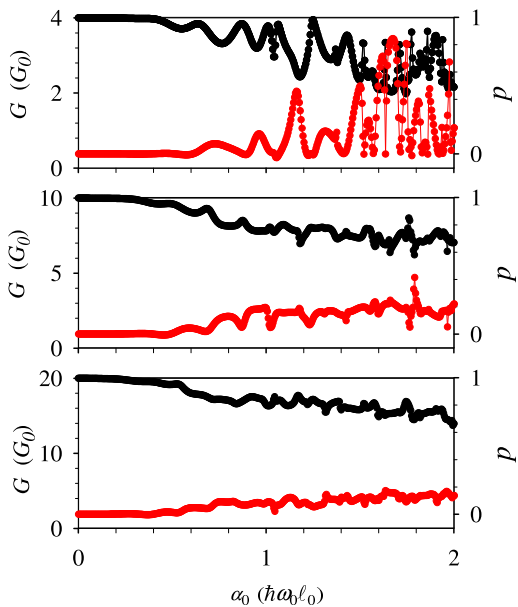,angle=0,width=0.39\textwidth,clip}
}
\caption{(Color online)
Conductance $G$, black symbols with left scale, and polarization of transmitted
current, grey symbols (red in color) with right scale, as a function of the Rashba
intensity. We have used the same parameters as in Fig.\ \ref{fig2}, except for
the Stoner fields which are here taken to vanish.
Upper, intermediate and lower panel correspond to
$N_p=4$, 10 and 20 propagating modes, respectively.
}
\label{fig6}
\end{figure}

In this section we assume nonmagnetic leads
by taking $\Delta_{\ell,r}=0$, i.e.,
vanishing Stoner fields in Fig.\ \ref{fig1}, and
analyze the evolution of the polarization and the conductance
when the number of active channels increases.
As shown in Fig.\ \ref{fig6} upper panel,
high polarizations $p$
are obtained for the minimal number of channels $N_p=4$
and strong spin-orbit intensities $\alpha_0$.
The clear correlation between $G$ and $p$, conductance minima
correspond to maxima in polarization, indicate that this is
an effect connected with the formation of quasibound states
that tend to block the current for a given spin
direction. When the number of channels is increased (lower panels
of Fig.\ \ref{fig6}) both $G$ and $p$ show
reduced staggering oscillations with increasing $\alpha$, as in
Figs.\ \ref{fig2}-\ref{fig5}. There is also an overall tendency
to smoothly reduce $G$ and increase $p$ in a linear way with $\alpha$.
With increasing number of channels the slopes of these
straight lines are reduced and
for $\alpha_0\approx2\hbar\omega_0\ell_0$ the polarization reaches
the values $\approx 0.2$ and $\approx 0.1$ for 10 and 20 propagating
channels, respectively.
In almost all cases the polarization is positive, indicating that
the transmitted current is preferentially polarized along $+y$.

\subsection{Smooth interfaces}

In this subsection we discuss how the results are affected by the
way in which the Rashba field is switched on spatially. For this, we vary
the parameter $a$ in the Fermi functions describing the transitions
shown in Fig.\ \ref{fig1}.\cite{Fermi} For large values of $a$ the edges
are quite smooth and correspond to an {\em adiabatic} turn-on
or turn-of in space. On the contrary, abrupt changes are given
by the limit $a\to 0$.
Our method is based on a grid discretization
of the variable $x$ and its only requirement is that the grid
should be fine enough to describe the spatial variations.

\begin{figure}[t]
\centerline{
\epsfig{file=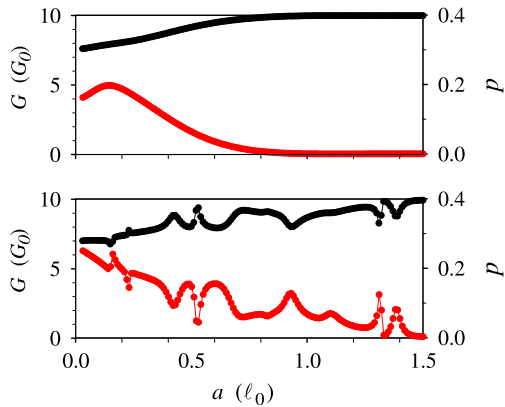,angle=0,width=0.40\textwidth,clip}
}
\caption{(Color online)
Conductance $G$, black symbols with left scale, and polarization of transmitted
current, grey symbols (red in color) with right scale, as a function of
the diffusivity $a$ in the Fermi functions describing the
spatial transitions in Fig.\ \ref{fig1}.
We have used the same parameters as in Fig.\ \ref{fig6},
and a value of the Rashba intensity
$\alpha_0=\hbar\omega_0\ell_0$ and $2\hbar\omega_0\ell_0$
for the upper and lower panels, respectively.
}
\label{fig7}
\end{figure}

The results discussed above have been obtained using
$a=0.1\ell_0$, a rather small value describing
abrupt transitions in space.
We have checked that either using a smaller value $a=0.05\ell_0$ or
a larger value $a=\ell_0$
the behaviors of the conductance in the presence of polarized leads discussed in Sec.\ II,
namely the staggering for high values of $\alpha_0$ and the modification
due to intersubband coupling, are not qualitatively changed.
Of course, it should be fulfilled that the Rashba dot length
$\ell$ is much greater than $a$ in order to still allow the
transition to reach to the saturation value $\alpha_0$.
More delicate is the polarization $p$ discussed in the preceding
subsection and Fig\ \ref{fig6}. In Fig\ \ref{fig7} we show the
evolution with $a$ of $G$ and $p$ when $N_p=5$ channels
are propagating in the wire. The polarization vanishes when
$a$ increases, indicating that smooth edges do not favor
the appearance of polarized currents. In this diffuse-edge
limit the conductance takes the maximal value $G=N_p G_0$ as
in a purely ballistic wire without any Rashba dot.
The evolution
for $\alpha_0=\hbar\omega_0\ell_0$ (upper panel)
is quite smooth but for $\alpha_0=2\hbar\omega_0\ell_0$
(lower panel) superimposed to the overall behavior we
find irregular maxima and minima as in previous results.

\section{Conclusions}

Recent experiments have proved the feasibility of the
spin transistor proposed by Datta and Das some years ago.\cite{Koo09,Dat90}
This device, usually presented as a paradigm of spintronics,
is expected to open new ways to overcome present limitations
of electronics. In this paper we have discussed some
specific aspects related to the Rashba interaction, including the so-called
intersubband coupling, relevant for a better understanding of the
physical mechanisms behind the spin transistors and spin polarizers.

Taking the wire containing the Rashba dot oriented along $x$ we have analyzed
the transmission in the presence of polarized leads along
$x$, $y$ or $z$,
and with increasing number of
propagating channels. The cases of parallel and antiparallel polarized leads
along $x$ and $y$ have been explicitly shown.
The evolution with Rashba intensity shows dramatic modifications when the
Rashba intersubband coupling is included. These modifications are specially
relevant at
strong values of $\alpha_0$, where staggering oscillations of
$G$ have been found.
In general, only a first smooth oscillation of
$G(\alpha_0)$ remains when the full Rashba interaction is considered,
while successive ones are heavily distorted or
even fully washed out. The spin-valve behavior is effectively destroyed by the
Rashba dot
and the conductance for both parallel and antiparallel leads
is relatively high.

The role of Rashba dots as spin polarizers has been discussed and
explicitly calculated assuming the leads to be nonpolarized. A smooth
linear increase in $p$ with Rashba intensity has been observed in the
multichannel case. In the limit of adiabatic transitions the polarization
vanishes. These overall smooth behaviors are superimposed by irregular
changes for high values of $\alpha_0$.

\section*{Acknowledgments}
Useful discussions with M.-S.\ Choi are gratefully acknowledged.
This work was supported by the MICINN (Spain) Grant FIS2008-00781.

\appendix

\section{Resolution method}
This appendix gives some details of the practical method to solve
Eq.\ (\ref{ccm}) and the
corresponding boundary conditions.
We use a method based on the quantum transmitting boundary algorithm.\cite{qtbm1,qtbm2}
A fictitious partitioning of the system in central and asymptotic
regions (contacts) is introduced. The boundaries for the left
and right contacts are at $x_\ell$ and $x_r$, respectively.
In the
contacts the band amplitudes take the form
\begin{equation}
\label{eccm1}
\psi_{ns}(x) =
a_{c,ns}\, e^{i s_c k_{c,ns} (x-x_c)}
+
b_{c,ns}\, e^{-i s_c k_{c,ns} (x-x_c)}
\;,
\end{equation}
where $c=\ell,r$ is a label referring to left $(\ell)$ and right $(r)$ contacts,
respectively, and we defined
$s_\ell=1$ and $s_r=-1$. The incident and reflected amplitudes for a given mode
$ns$ and contact $c$ are given by $a_{c,ns}$ and $b_{c,ns}$, respectively.
This expression is for a propagating channel in contact $c$, for which
$\varepsilon_n+|\Delta_c|+s\Delta_c<E$
and its corresponding wavenumber
\begin{equation}
k_{c,ns}=\sqrt{2m^*(E-\varepsilon_n-|\Delta_c|-s\Delta_c)}/\hbar\; ,
\end{equation}
is a real number.
Equation (\ref{eccm1}) also applies to evanescent modes,
$\varepsilon_n+|\Delta_c|+s\Delta_c>E$,
if we assume in this case $a_{c,ns}=0$ and a purely imaginary wavenumber
\begin{equation}
k_{c,ns}=i\sqrt{2m^*(\varepsilon_n+|\Delta_c|+s\Delta_c-E)}/\hbar\;.
\end{equation}
Notice that the output amplitudes can be obtained from the wave function right at
the interface,
\begin{equation}
\label{eqbcn}
b_{c,ns} = \psi_{ns}(x_c) - a_{c,ns}\; .
\end{equation}
Substituting Eq.\ (\ref{eqbcn}) in Eq.\ (\ref{eccm1}) we obtain
\begin{eqnarray}
\label{eccm2}
\psi_{ns}(x)-\psi_{ns}(x_c)\,e^{-i s_c k_{c,ns} (x-x_c)}
=&&\nonumber\\
2 i a_{c,ns}  \sin(s_c k_{c,ns}&&\!\!\!\!\!\!\!
(x-x_c))\,,
\end{eqnarray}
that is the quantum-transmitting-boundary equation for the contacts.

Equations (\ref{ccm}) and (\ref{eccm2}), for the central and contact regions,
respectively, form a closed set that does not invoke the wave function
at any external point. Of course, this is not true for any of these two subsets separately,
since central and contact regions are connected
through the derivative in Eq.\ (\ref{ccm}) and of $\psi_{ns}(x_c)$ in Eq.\ (\ref{eccm2}).
In practice, we use a uniform grid in $x$ with $n$-point formulae for the derivatives ($n\approx5-11$)
and truncate the expansion in transverse bands, Eq.\ (\ref{exp}), to include typically 30-60 terms.
The resulting sparse linear problem is then solved using
routine \verb+ME48+.\cite{harwell}


\begin{thebibliography}{90}
\bibitem{Fermi}
The inhomogeneous Rahba intensity is given by $\alpha(x)=\alpha_0[F(x,\ell/2)-F(x,-\ell/2)]$ where
the Fermi functions are $F(x,x_0)=[1+\exp((x-x_0)/a)]^{-1}$ and $a$ is a parameter
modeling the diffusivity
of the interface. Similar expressions for the gate potential $V_g(x)$ and Stoner field $\Delta(x)$
are used.
\end{thebibliography}
\end{document}